\documentclass[showpacs,aps,epsfig,nofootinbib]{revtex4}
\usepackage{mathrsfs}

\usepackage{graphicx}
\usepackage{amsfonts}
\usepackage{epstopdf}
\usepackage{latexsym}
\usepackage{amssymb}
\usepackage{amssymb}
\usepackage{ulem}
 \usepackage{color}

\usepackage[center]{subfigure}
\begin{document}

 \newcommand{\bq}{\begin{equation}}
 \newcommand{\eq}{\end{equation}}
 \newcommand{\bqn}{\begin{eqnarray}}
 \newcommand{\eqn}{\end{eqnarray}}
 \newcommand{\nb}{\nonumber}
 \newcommand{\lb}{\label}
\newcommand{\PRL}{Phys. Rev. Lett.}
\newcommand{\PL}{Phys. Lett.}
\newcommand{\PR}{Phys. Rev.}
\newcommand{\CQG}{Class. Quantum Grav.}


\title{Applying deep neural networks to the detection and space parameter estimation of compact binary coalescence with a network of gravitational wave detectors}

\author{Xilong Fan}
\affiliation {School of Physics and Technology, Wuhan University, Wuhan 430072, China}
\author{Jin Li}
\email{cqujinli1983@cqu.edu.cn (corresponding-author)} 
\affiliation {Department of Physics, Chongqing University,
Chongqing 401331, China}
\author{Xin Li}
\affiliation {Department of Physics, Chongqing University, chongqing 401331, China}
\author{Yuanhong Zhong}
\affiliation {College of communication, Chongqing University,
Chongqing 400044, China}
\author{Junwei Cao}
\affiliation {Research Institute of Information Technology, Tsinghua University, Beijing 100084, China}

\date{\today}

\begin{abstract} 
In this paper, we study an application of deep learning to the advanced LIGO and advanced Virgo coincident detection of gravitational waves (GWs) from compact binary star mergers. This deep learning method is an extension of the Deep Filtering method used by George and Huerta (2017) for multi-inputs of network detectors. Simulated coincident time series data sets in advanced LIGO and advanced Virgo detectors are analyzed for estimating source luminosity distance and sky location. As a classifier, our deep neural network (DNN) can effectively recognize the presence of GW signals when the optimal signal-to-noise ratio (SNR) of network detectors $\geq$ 9. As a predictor, it can also effectively estimate the corresponding source space parameters, including the luminosity distance $D$, right ascension $\alpha$, and declination $\delta$ of the compact binary star mergers. When the SNR of the network detectors is greater than 8, their relative errors are all less than 23\%. Our results demonstrate that Deep Filtering can process coincident GW time series inputs and perform effective classification and multiple space parameter estimation. Furthermore, we compare the results obtained from one, two, and three network detectors; these results reveal that a larger number of network detectors results in a better source location. 

\end{abstract}


\maketitle

\section{Introduction}
 As the authority of gravitational wave (GW) detection, last year, the advanced Laser Interferometer Gravitational-Wave Observatory (advanced LIGO) and advanced Virgo performed the first three-detector detection of a GW signal from a binary neutron star coalescence \cite{prlnew}. This is the fifth direct detection of GWs from compact binary coalescences by advanced LIGO \cite{prl1,prl2,prl3,prl4,add3}. With the continuous improvement of advanced LIGO and advanced Virgo \cite{AdLIGO, AdVirgo,add1,add2}, numerous GW triggers are expected to appear in the time series. Since, the current data analysis of advanced LIGO and advanced Virgo is computationally expensive for detecting the matched GW signals in noisy data and for distinguishing signals from glitches, new methodologies for signal processing are required that can process data rapidly and with high accuracy. In recent years, the application of machine learning to GW detection has been widely proposed \cite{MLGW1,MLGW2,MLGWadd,MLGW3,MLGW4,MLGW5,new17,new18,new19,new20}. Machine learning enables a computer to use specific learning algorithms to program itself based on large supplies of data to solve particular problems \cite{MLoverview}. Among them, a deep learning architecture that is usually in the form of deep neural networks (DNNs) simulates the learning process of a human brain. By processing input information and sharing it with relevant neurons with sufficient neural connections, deep learning enables autonomous learning from data, yielding simpler data analysis \cite{rep17,rep18}. Most importantly, its ability to directly process raw noisy time series has been demonstrated, which can enable real-time GW observation \cite{MLGW2}. Deep learning is expected to be a popular approach for processing big data and analyzing problems in the field of astrophysics. 

Traditional NNs composed of three layers---the input, hidden, and output layers---have a limited ability to solve problems. However, researchers have discovered that if the hidden layer is extended with additional layers, its solving ability can be greatly enhanced \cite{DNN1}. NNs with long, interconnected layers between the input and output layers are called deep neural networks (DNN). Depending on the properties of the task, many categories of DNNs can be used, such as a recurrent neural network (RNN) \cite{DNN1,RNN1, MLGW2} for exhibiting temporal behavior in data with variable input lengths, and a convolutional neural network (CNN) \cite{UIUC66,UIUC67,UIUC68} for extracting features from data. Because a CNN is much easier to train due to its local receptive fields and weight sharing, and its effectiveness in automatically recognizing input, we have selected CNN as an integral part of our DNN system. Comparing the DNN to the matched filtering, widely used in the advanced LIGO's and advanced Virgo's pipelines, indicates that the DNN can significantly quicken GW searches \cite{MLGW2}.
 
There are a variety of GW sources in the advanced LIGO's and advanced Virgo's frequency band, such as the mergers of compact binary stars, which have been researched intensively. In particular, binary neutron stars have proven to be associated with abundant electromagnetic counterparts \cite{BNS1,BNS2}. As in \cite{MLGW2}, we first train the DNN as a classifier to distinguish signal from noise. Second, the DNN structure is adjusted and trained as a predictor to estimate a source's space parameters, which can constrain the locations of compact binary stars and provide insight for future multi-messenger observations \cite{Mmess1,Mmess2,Mmess3,Mmess4,Fan}.  

In this paper, we take into account the coincident detection of three detectors, H, L, and V (HLV) referring to the advanced LIGO-Hanford, advanced LIGO-Livingston, and advanced Virgo detectors, respectively, and perform multiple space parameters estimation directly from noisy time series inputs after our DNN is trained by simulated signals. Gabbard et al. \cite{new19} has shown that a deep convolutional neural network can reproduce the sensitivity of a matched-filtering search for simulated binary black hole gravitational-wave signals. In the first application of the Deep Filtering method to advanced LIGO and advanced Virgo GW detection, George and Huerta have detected GW150914 from noisy time series in one detector and effectively measured their masses \cite{MLGWadd}. We, then, study the DNN as a predictor to estimate the luminosity distance and sky location of binary black holes. This paper is organized as follows. In section II, we describe the theoretical model for building our training and test data sets. In section III, we introduce the principle of DNN algorithms and investigate the performance of our designed DNN as a classifier and predictor, respectively. Conclusions and remarks are presented in section IV.

\section{Obtaining training and test data}
The GW signal from a compact binary coalescence can be divided into three phases: the inspiral, merger, and ringdown of the final object. The waveform $h_{+}$, $h_{\times}$ of all the phases has been successfully simulated using the effective-one-body model. Herein, we adopt the \text{"EOBNRv4"} to generate training and test data sets through the PyCBC software package (https://ligo-cbc.github.io). The GW strain $h(t)$ of an interferometer is a linear combination of $h_{+}$ and $h_{\times}$: 
\begin{equation}
h(t)= F^{+}(t)h_{+}(t)+F^{\times}(t)h_{\times}(t),
\end{equation}
where, the antenna functions $F^{+}(t)$ and $F^{\times}(t)$ for a specific
detector geometry are mainly related to the angles describing GW polarization $\psi$ and source location at a celestial sphere frame coordinate (right ascension $\alpha$, declination $\delta$), and the positions and orientations of the detectors. The detailed expressions of these functions, as well as the specific latitude of the detector's location along with the orientation of each of the detector's arms with respect to the local geographical directions, can be found in \cite{addfan,detectorlocation}.

\begin{figure}
\includegraphics[height=8cm]{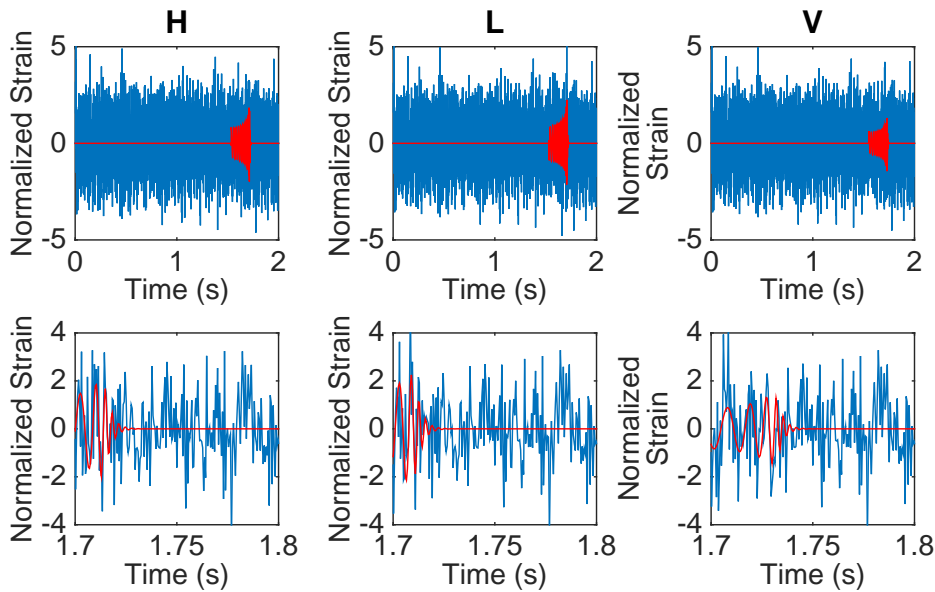}
\caption{Top panel: Blue curves are simulated time series in the advanced LIGO and advanced Virgo detectors, respectively, corresponding to the similar GW event and inputted to the DNN together. Bottom panel: Zoom plots corresponding to the top plots around injection time, which varies among detectors owing to their different locations \cite{addfan, networksGWreviewlatest}. The sample frequency is $f_{s}=2048\text{Hz}$.}\label{Expected-Signal}

\end{figure}

\begin{figure}
\centerline{\includegraphics[height=5cm]{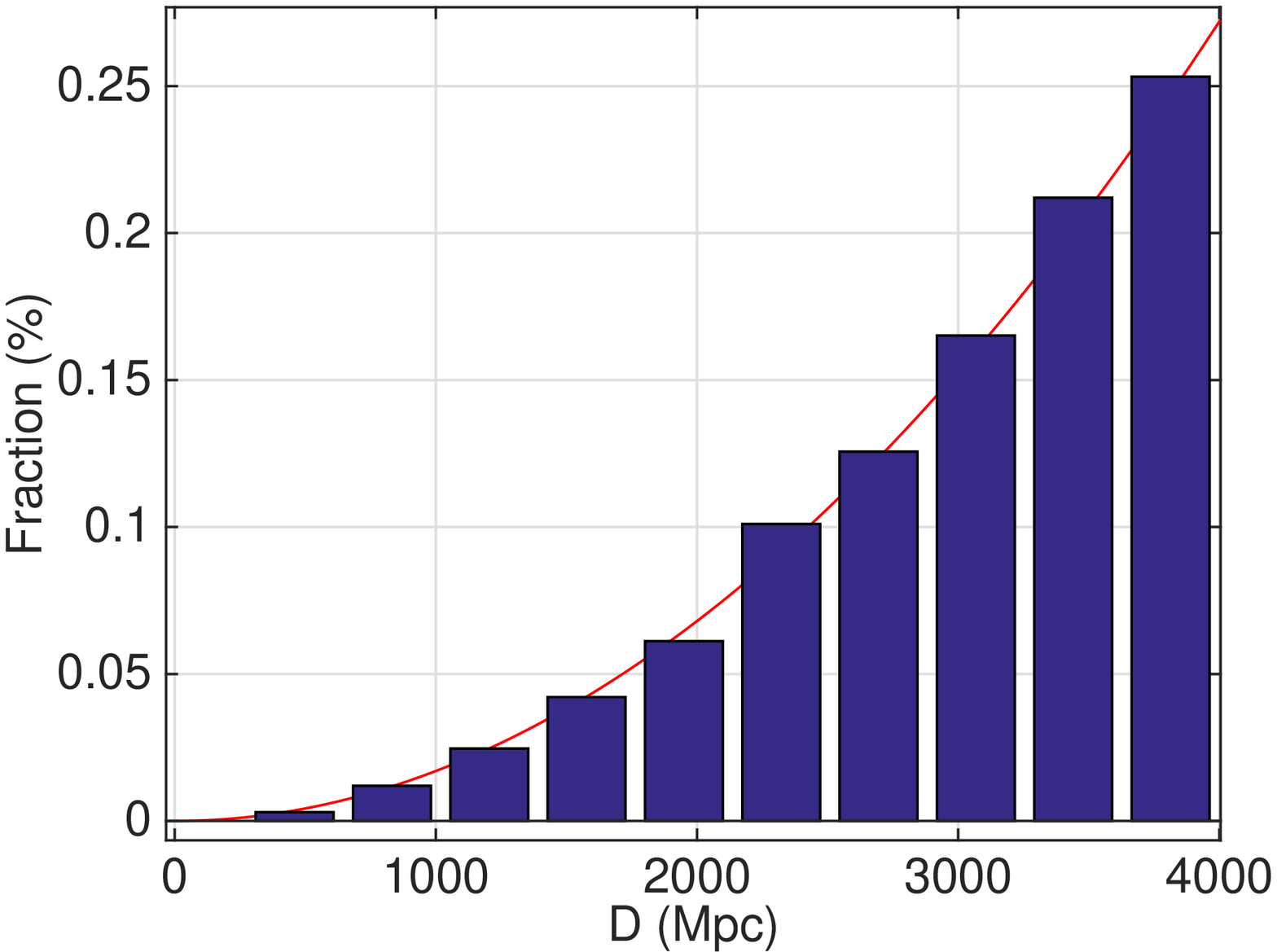}~~~~~~\includegraphics[height=5.5cm]{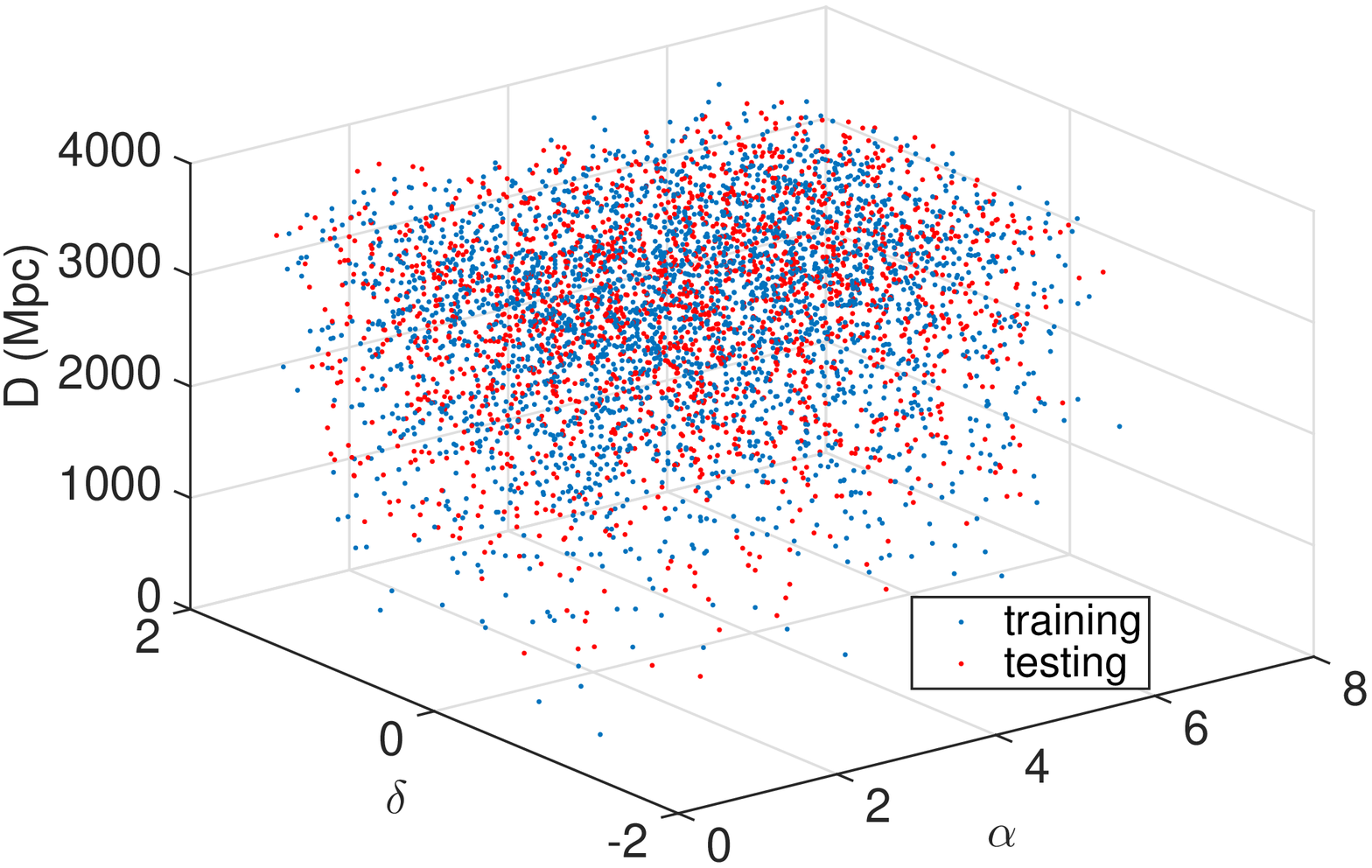}}
\caption{Left panel: Distribution of the source luminosity distance $D$. Right panel: Source luminosity distance $D$ and angles at the celestial sphere frame coordinate $(\alpha,\delta)$ chosen for the training and test data sets. Here, the optimal SNR$ is 9$. }\label{training--testing}

\end{figure}

GW signals are submerged in Gaussian noise with the advanced LIGO's and advanced Virgo's noise power spectral density (PSD) at designated sensitivities, respectively \cite{HLsensitivity,Vsensitivity}. Fig. \ref{Expected-Signal} shows one sample of a coincident GW signal in the three detectors. The prior distributions on $\alpha,\delta,\psi$ and the orbital inclination $\iota$ in $h_{+}, h_{\times}$ should be uniformly distributed in their respective ranges. The priors on $m_1$, $m_2$ are assumed to be uniform in the range from 10 to 40 M$_{\odot}$, and the distribution of luminosity distance $D$ are assumed to be uniform on the sky spherical surface from 10 to 4000 Mpc (see Fig. \ref{training--testing}). In each time series, all of our simulated binary black hole mergers are included in a 2-second window with cutoff frequency $f_{\rm{cutoff}}=40\rm{Hz}$, and the arrival time of the GW at the Earth's center is selected from the uniform distribution on the interval [1.7, 1.75]s. For each detector, the time strain is 
\begin{equation}
s_{i}(t_{0}+\tau_{i}+t)= F_{i}^{+}(t)h_{+}(t)+F_{i}^{\times}(t)h_{\times}(t)+n_{i}(t_{0}+\tau_{i}+t),~~~~0<t<T,
\end{equation}
where, $t_{0}$ is the time at which the GW reaches the origin of any fixed coordinate, $\tau_{i}$ is the time at which the GW propagates from the origin to i$^{th}$ detector, and $T$ is the duration of the GW. The time delay and antenna functions for each detector are based on the space locations of the advanced LIGO and the advanced Virgo (cf. Table 1 of \cite{networksGWreviewlatest}). In this paper, we adopt the optimal SNR $\rho=2[\int_{0}^{\infty}|\tilde{h}(f)|^{2}df/S_{h}(f)]^{1/2}$ to describe the SNR \cite{GWrevpaper}, where, $S_{h}(f)$ is the noise PSD. In the following calculations, all of the SNRs are calculated from network detectors.

Splitting data into separate sets for training and testing is necessary for supervised learning. We randomly generate 3,000 training sets and 2,000 test sets with a given SNR (see right panel of Fig. \ref{training--testing}). In addition to these, 2,000 validation sets are separately generated for each SNR in order to adjust the hyperparameters of the DNN. We, then, standardize all of the sets. Herein, we select a sampling frequency of $2048\text{Hz}$ (cf. Fig. 6 of \cite{wavelet}). For coincident observation, the data from H, L, and V should be inputted together, leading to a 3$\times$4096 dimensional tensor for each input data set. In total, we generate 100,000 sets for training by adding different noise to 10,000 templates, and 2,000 sets for testing with each specific SNR. 

\section{DNN for GW detection and space parameter estimation}
\begin{figure}
\centerline{\includegraphics[height=5cm]{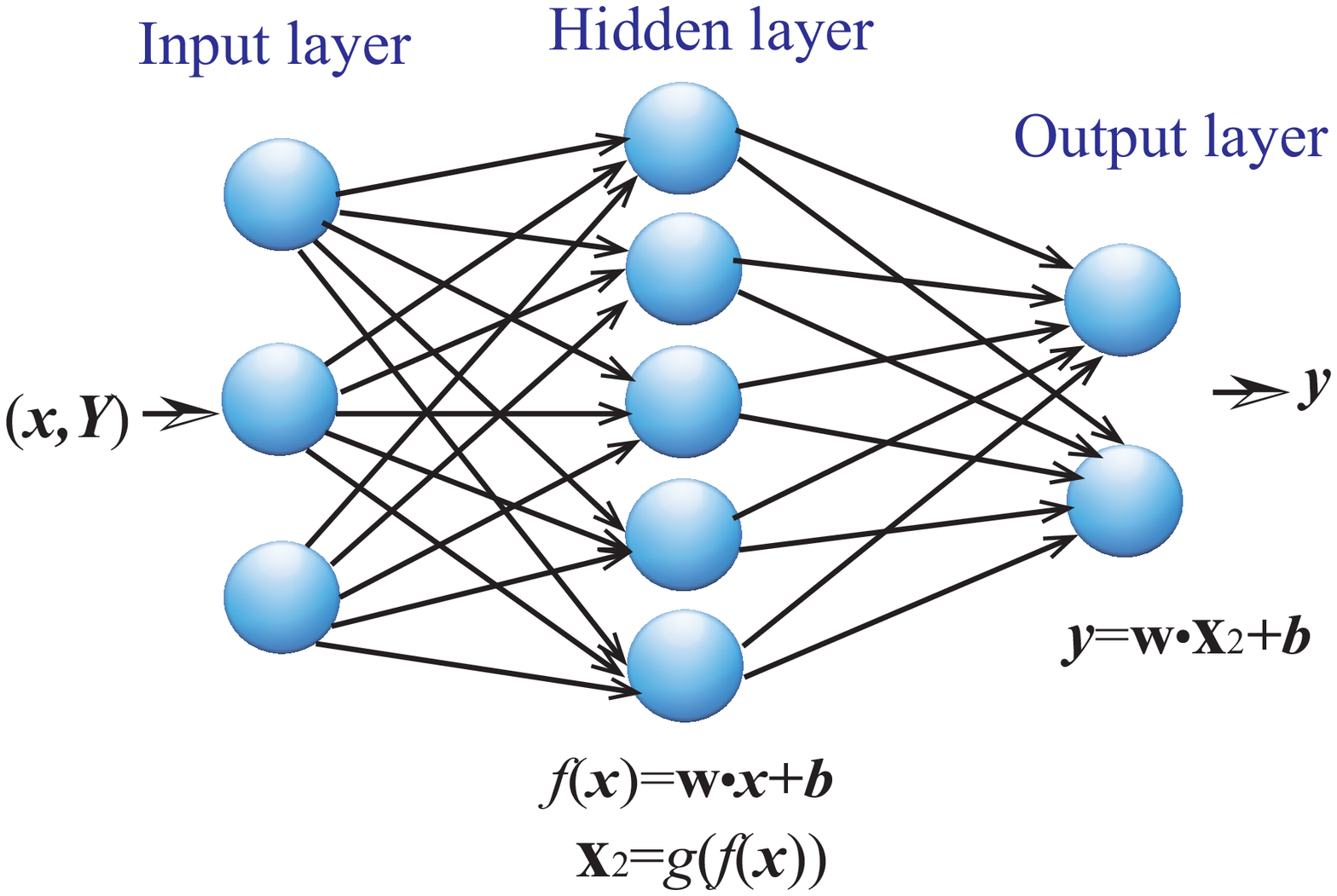}~~~~~~\includegraphics[height=5cm]{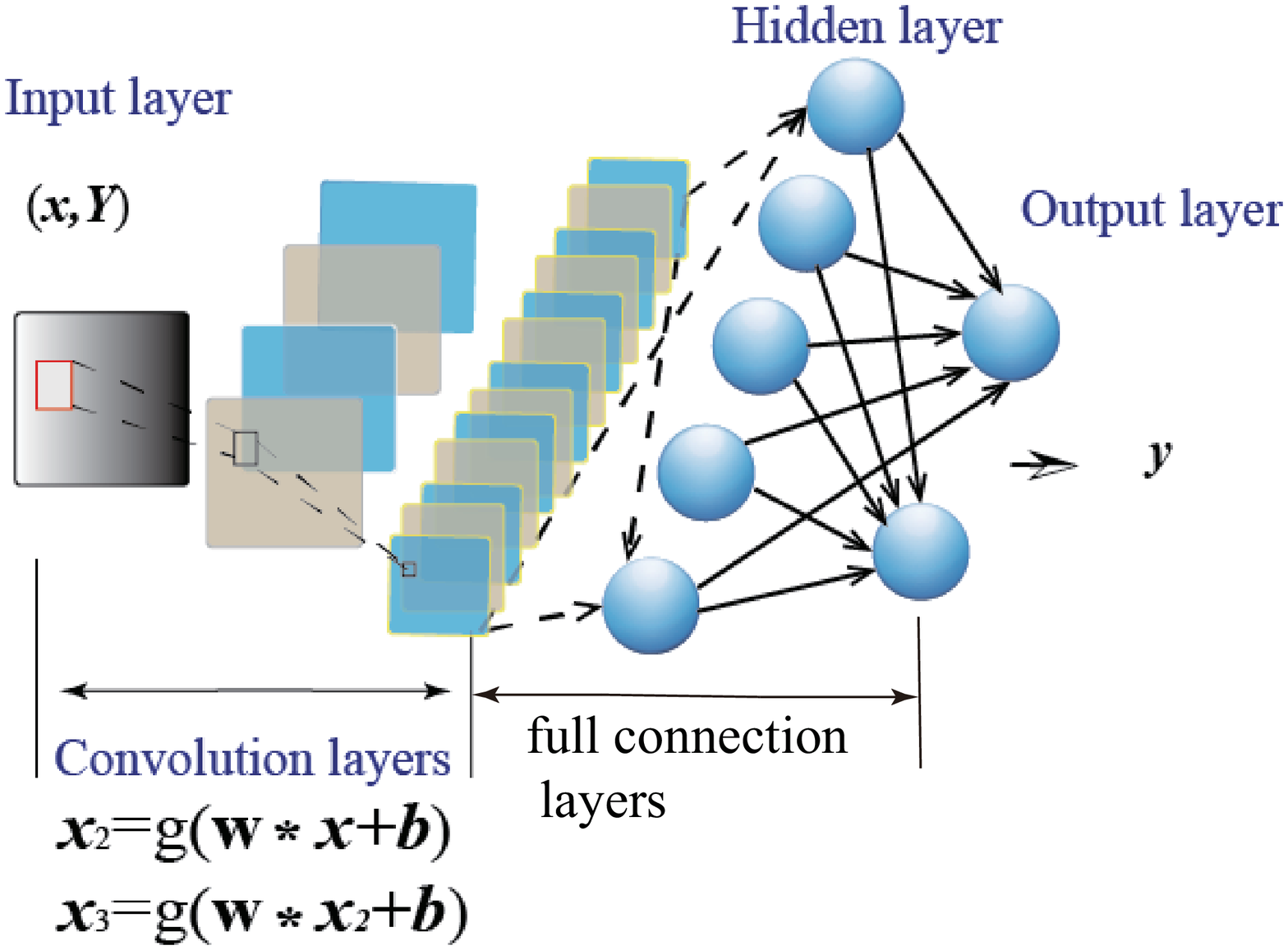}}
 \caption{Left panel: Traditional neural network (NN), in which the neurons (blue spheres) are connected to neurons in other layers by synapses (arrows). In the training process, the NN gets input $\vec{x}$ and $\vec{Y}$ and produces output $\vec{y}$. Right panel: Illustration of a DNN structure with convolution layers and fully connected layers, where $*$ is a convolution operator.}\label{fig:NN}
\end{figure}

{\bf Structure of a DNN:} An NN composed of many neurons can simulate the learning process of a human brain and is one of the most popular methods in machine learning. The standard structure of an NN includes an input layer, a hidden layer, and an output layer (see left panel of Fig. \ref{fig:NN}). Information enters the input layer and is transferred to neurons in the next layer (i.e., the hidden layer) via synapses. Using vector $\vec{x}$ as an example, herein, the input information is expressed by linear combinations as $f(\vec{x})=\vec{w}\cdot\vec{x}+b$, where $\vec{w}$ and $b$ are weights and bias, respectively \cite{MLGW2}, and $\vec{Y}$ is the real answer to $\vec{x}$. The weights and bias are learned through training. In the hidden layer, the information is split; thus, it becomes simpler to analyze. To obtain a desired result, a regression analysis is applied on the synapses between the hidden and output layers; for instance, a logistic regression analysis is used for classification. Based on the features of the learning problem, certain activation functions $g(x)$, such as the logistic sigmoid, hyperbolic tan, and rectified linear unit (ReLU or Ramp) \cite{MLGW2}, are applied to the output of the hidden layer, i.e., $\vec{y}=g(f(\vec{x}))$. The output layer is a logistic regression layer containing two neurons as a classifier corresponding to two classes (i.e., true or false) and a number of neurons as a predictor corresponding to the number of estimated space parameters. Regarding the mathematical operations, some references provide detailed explanations \cite{CNN1}. Occasionally, there is more than one hidden layer. Because each neuron branches out to connect with all the neurons in the adjacent layer, such an NN is called a fully connected neural network and correspondingly the layers are called fully connected layers.  
 
 \begin{table}[h]
\centering
\caption{Modified version of the DNN in \cite{MLGW2} used for classification; it is also used for prediction by replacing the 15$^{\text{th}}$ layer with a ramp (ReLU) function (same method as \cite{MLGW2}) and changing the size of layer 14 according to the dimension of predicted parameters. The number of neurons in the consecutive layers is 8, 16, and 32, respectively. The kernel sizes for the convolutional layers are 3$\times$16, 1$\times$16, 1$\times$16, and 1 for all the pooling layers. The stride is set to 1 for all of the convolutional layers and 4 for all of the pooling layers. The dilation factor of the convolutional layers are set to 1, and the padding size of the pooling layer is zero. The function of pooling is max. }\label{fig:classifier}
\begin{tabular}{|c|c|}

\hline  
Input&Matrix(3$\times$4096)\\
\hline  
1~ReshapeLayer~~~~~~&3-tensor(size:1$\times$3$\times$4096)\\
2~ConvolutionLayer&3-tensor(size:8$\times$1$\times$4081)\\
3~PoolingLayer~~~~~~~&3-tensor(size:8$\times$1$\times$1021)\\
4~Ramp~~~~~~~~~~~~~~~~&3-tensor(size:8$\times$1$\times$1021)\\
5~ConvolutionLayer&3-tensor(size:16$\times$1$\times$1006)\\
6~PoolingLayer~~~~~~~&3-tensor(size:16$\times$1$\times$252)\\
7~Ramp~~~~~~~~~~~~~~~~&3-tensor(size:16$\times$1$\times$252)\\
8~ConvolutionLayer&3-tensor(size:32$\times$1$\times$237)\\
9~PoolingLayer~~~~~~~&3-tensor(size:32$\times$1$\times$60)\\
10~Ramp~~~~~~~~~~~~~~~~&3-tensor(size:32$\times$1$\times$60)\\
11~FlattenLayer~~~~~~~&vector(size: 1920)\\
12~LinearLayer~~~~~~~~&vector(size: 64)\\
13~Ramp~~~~~~~~~~~~~~~~&vector(size: 64)\\
14~LinearLayer~~~~~~~~&vector(size: 2)\\
15~SoftmaxLayer~~~~~~&vector(size: 2)\\
~~~~~~Output~~~~~~~~~~&class\\
\hline 
\end{tabular}
\end{table}

Commonly, some practical learning problems need more than one hidden layer. If a fully connected network is used, a large number of hyperparameters must be determined including the number of hidden layers and neurons in each layer, the choice of activation functions, the learning rate, and the iterations, and it is very difficult to train such a network. Convolutional neural network (CNNs) have been explored for the neocognitron \cite{UIUC66, MLGW2}, a neural network that is not fully connected, since each neuron is connected to only a few neurons in the following layer, and that shares weights among the neurons. By performing a convolution calculation, a CNN can learn signal features by rotating and scaling signals. Due to their effective feature extraction and avoidance of redundant hyperparameter calculations, CNNs have obtained remarkable success in the fields of computer vision \cite{cv} and natural language processing \cite{lp}. Currently, the application of CNNs to time series data processing is being developed \cite{timeCNN, MLGW2}. In fact, DNNs, which combine fully connected layers with convolutional and modification layers, have been proposed to be more effective for both signal detection and parameter estimation from noisy time series data \cite{MLGW2}. The right panel of Fig. {\ref{fig:NN}} provides an illustration of the DNN structure. In the convolution layers, output from the previous layer, $\vec{x}_{k}$, is filtered by neurons through convolution operations and transferred to the output of this layer, $\vec{x}_{k+1}$, (i.e., $\vec{x}_{k+1}=g(\vec{w}*\vec{x}_{k}+b)$, where, $*$ is the convolution operator). Therefore, we plan to use this type of DNN to learn how to recognize GW signals from raw time series data sets and estimate the corresponding parameters of the GW sources. Determining the optimal hyperparameters of the DNN, including the number of convolutional and fully connected layers, size of pooling layers, number of neurons in each layer, learning rate, iterations, and activation functions, remains a challenging problem \cite{DNN1}. Currently, several approaches for determining optimal hyperparameters have been proposed, such as the randomized trial-and-error-based methods adopted in this paper(http://neuralnetworksanddeeplearning.com), Bayesian optimization \cite{UIUC102}, and genetic algorithms \cite{GenCNN}. 

{\bf Training process:} Once the structure of a DNN is determined, then, for each attempted hyperparameter, the DNN should be trained. As in human learning, the training process is the most important step for DNN learning, and the learning algorithm plays a key role in this process. The goal of training is to determine the optimal weights and bias of the DNN. Initially, the computer provides random weights, $\vec{w}$, and bias, b, to the DNN, which then undergoes several rounds of training using the training input data and obtains output $\vec{y}$. Finally, the optimal $\vec{w}$ and b are determined when the loss function $l(\vec{w},b)$, which describes the difference between the DNN result $\vec{y}$ and the real answer $\vec{Y}$, reaches the minimum value. The loss function differs for classification and parameter estimation, which is discussed further in the following subsections. Applying each trained DNN to the validation sets, the optimal hyperparameters can be determined according to accuracy. Our optimal hyperparameters are illustrated in Table \ref{fig:classifier}.

In this paper, we adopt the ADAM method, which uses stochastic gradient descent with an adaptive rate, to train the DNN \cite{ADAM}. Gradient descent with an adaptive rate is invariant to the diagonal rescaling of the gradients of the loss function. When the local minimum is reached, the first-order partial differential equations equals to zero: $\partial l/\partial \vec{w}=\partial l/\partial b=0$. This applies only to neurons from the input layer to the output layer; for adjusting the weights of neurons in the preceding layers, a back-propagation algorithm should be used \cite{rep24}. From the updated weights in the output layer and by differentiating the activation function, back-propagation can be used to determine the updated weights in all preceding layers \cite{BP1}. On the Mathematica 11.1 platform, we use the $NetChain$ command to design the DNN, whose hyperparameters are selected by random trials. Then, for each designed DNN, we apply the $NetInitialize$ command to initialize the network with a random distribution of weights and bias before training it.

\subsection{DNN as a classifier}
The main task of the DNN as a classifier is to detect GW signals in numerous time series data sets. The aim of the classifier is to correctly distinguish the class corresponding to $\vec{x}$ with maximum likelihood; therefore, the loss function is defined as the negative mean log-likelihood: 
\begin{equation}
l(N, \vec{Y}, \vec{x})=-\frac{1}{N}\left[\sum_{i=1}^{N}\text{log}(P(\vec{Y}=\vec{y}|x_{i}))\right],
\end{equation}
where, $x_{i}$ represents the $i^{th}$ training input data set and $N$ is the size of the training data sets, which in our case is 100,000. $\vec{Y}$ and $\vec{y}$ are the real and predicted classes for $x_{i}$, respectively. $\text{log}P(\vec{Y}=\vec{y}|x_{i})$ represents the log probability of the predicted class of the DNN, $\vec{y}$, is the real class $\vec{Y}$ for a given $x_{i}$. 
 
The input of the classifier comprises several 3$\times$4096 time series from three detectors. Each sample is recorded as three time series, which are simultaneously measured by three separate detectors, H, L, and V. Based on the structure in \cite{MLGW2}, we have made some modifications. First, the three time series inputs from H, L, and V are joined into a 1$\times$12288 matrix using a reshape layer, since the dimension of each neuron can then be simplified without any loss in accuracy in feature extraction. Second, specific values for kernel size, stride, dilation, and other DNN hyperparameters are set. Third, the DNN is trained. Then, validation sets are used to check the effectiveness of the trained DNN. After selecting a wide variety of hyperparameters for the DNN, the structure of our classifier is finally determined (shown in Table \ref{fig:classifier}) based on the accuracy and operation time of the hyperparameters. Later on, we use NVIDIA GPU (Gtx1080ti, 11G ram) to implement the deep learning process.

\begin{figure}
\includegraphics[height=5cm]{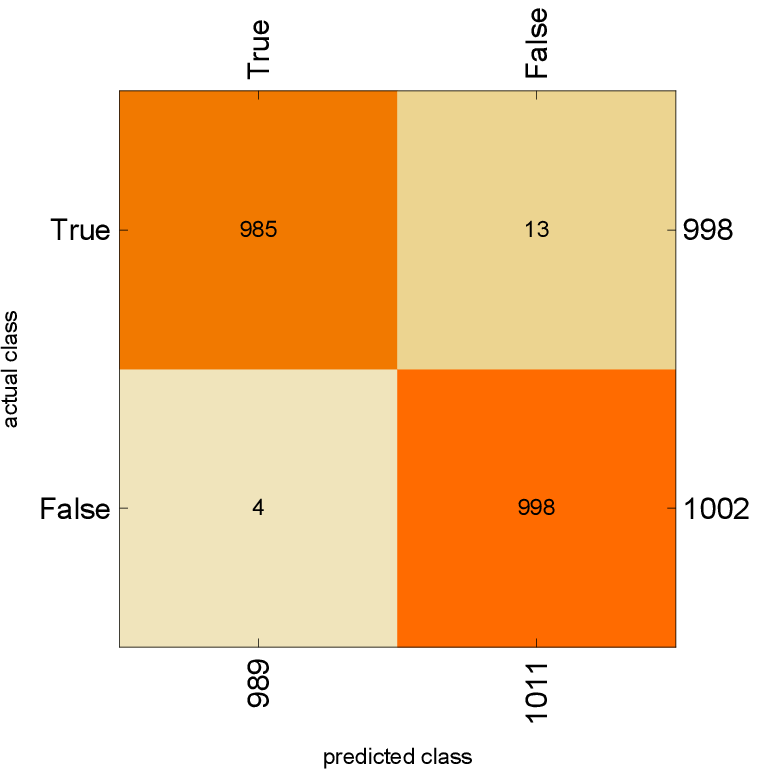}\includegraphics[height=5cm]{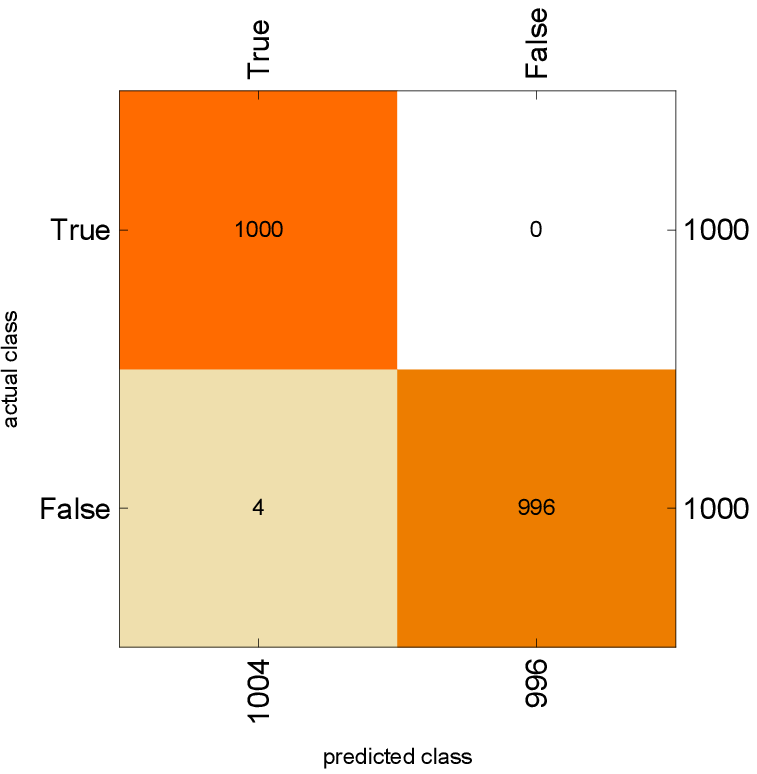}\includegraphics[height=4.5cm]{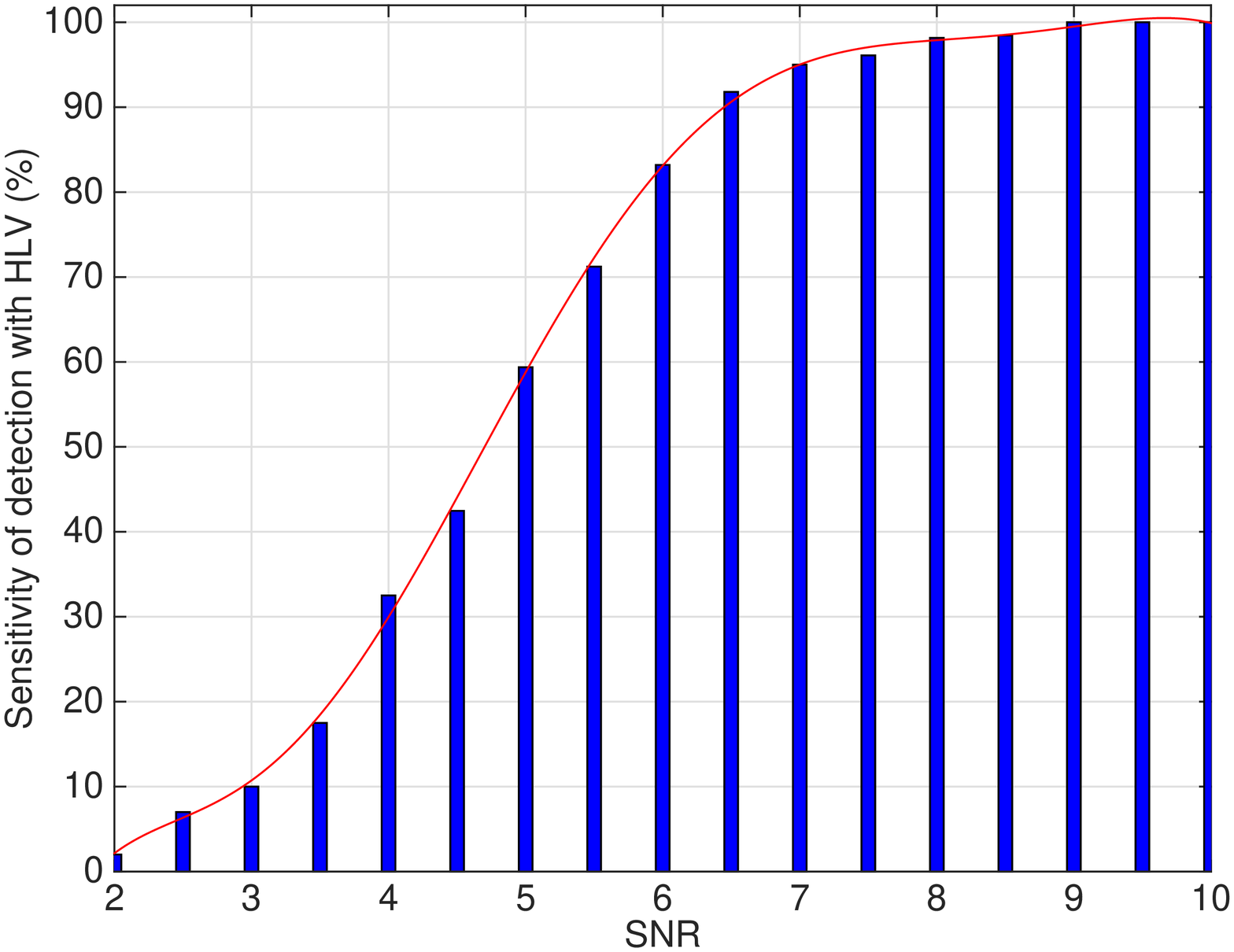}
 \caption{Left panel: confusion matrix of a classifier on a test set with SNR = 8 (sensitivity is 98.70$\%$). Middle panel: confusion matrix of a classifier on a test set with SNR $=9$ (sensitivity is 100$\%$ for all signals with a higher SNR). Right panel: sensitivity of detection varying with the SNR, where, the false alarm rate $=0.4\%$.}\label{fig:acy_classifier}
\end{figure}

After training for nearly nine hours, the classifier obtains 98.70$\%$ sensitivity of detection for SNR = 8 and 100$\%$ sensitivity for SNR $\geq$9, with a false alarm rate of 0.4\%. The confusion matrices are shown in the left and middle panels of Fig. \ref{fig:acy_classifier}. The right panel of Fig. \ref{fig:acy_classifier} illustrates the sensitivity varying with the SNR in the three network detectors, where, sensitivity increases for a higher SNR.

\subsection{DNN as a predictor}
With a structure similar to that of our classifier, the predictor uses a ReLU function instead of the final softmax layer \cite{MLGW2} and a different output size. When trained with a template bank of expected signals, a DNN is able to estimate multiple parameters directly from noisy time series data \cite{MLGW2}. Herein, we focus on the parameters $D$, $\alpha$, and $\delta$ for multiple parameter estimation, which directly provide the location of sources. With respect to the DNN as a predictor, we also compare its performance on one, two, and three network detectors, which can reveal meaningful insights into the coincident observation of GWs. 

For parameter estimation, the loss function is generally the mean squared error:
\begin{equation}
l(N,\vec{Y},\vec{y})=\frac{1}{N}\left(\sum_{i=1}^{N}{(Y_{i}-y_{i})}\right)^{2},
\end{equation}
where, $Y_{i}$ and $y_{i}$ are the real and predicted values of parameters $D_{i}, \alpha_{i}, \delta_{i}$, respectively, corresponding to the $i^{th}$ training data set, and $N$ is the number of training data sets.

\begin{figure}
\includegraphics[height=4cm]{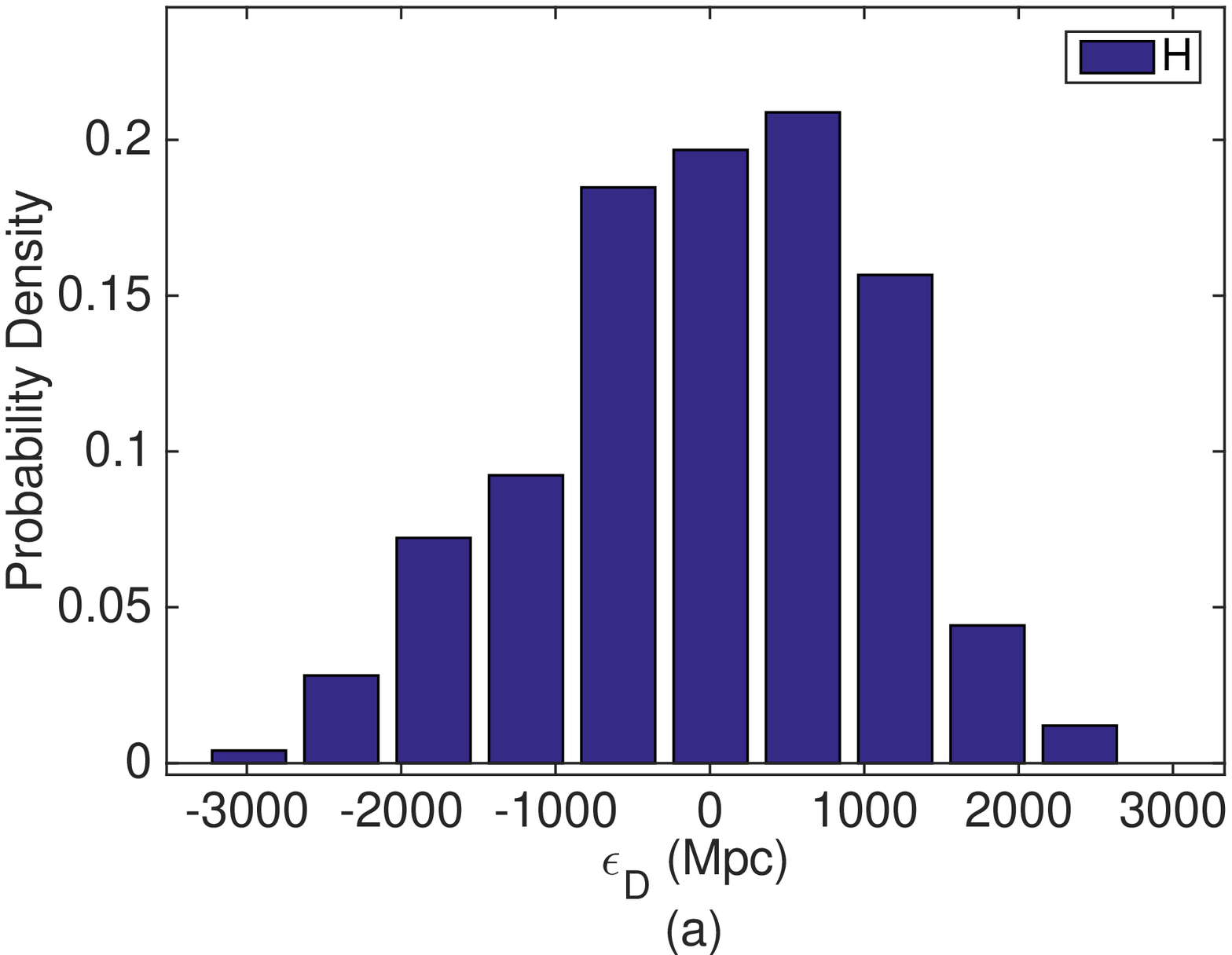}\includegraphics[height=4cm]{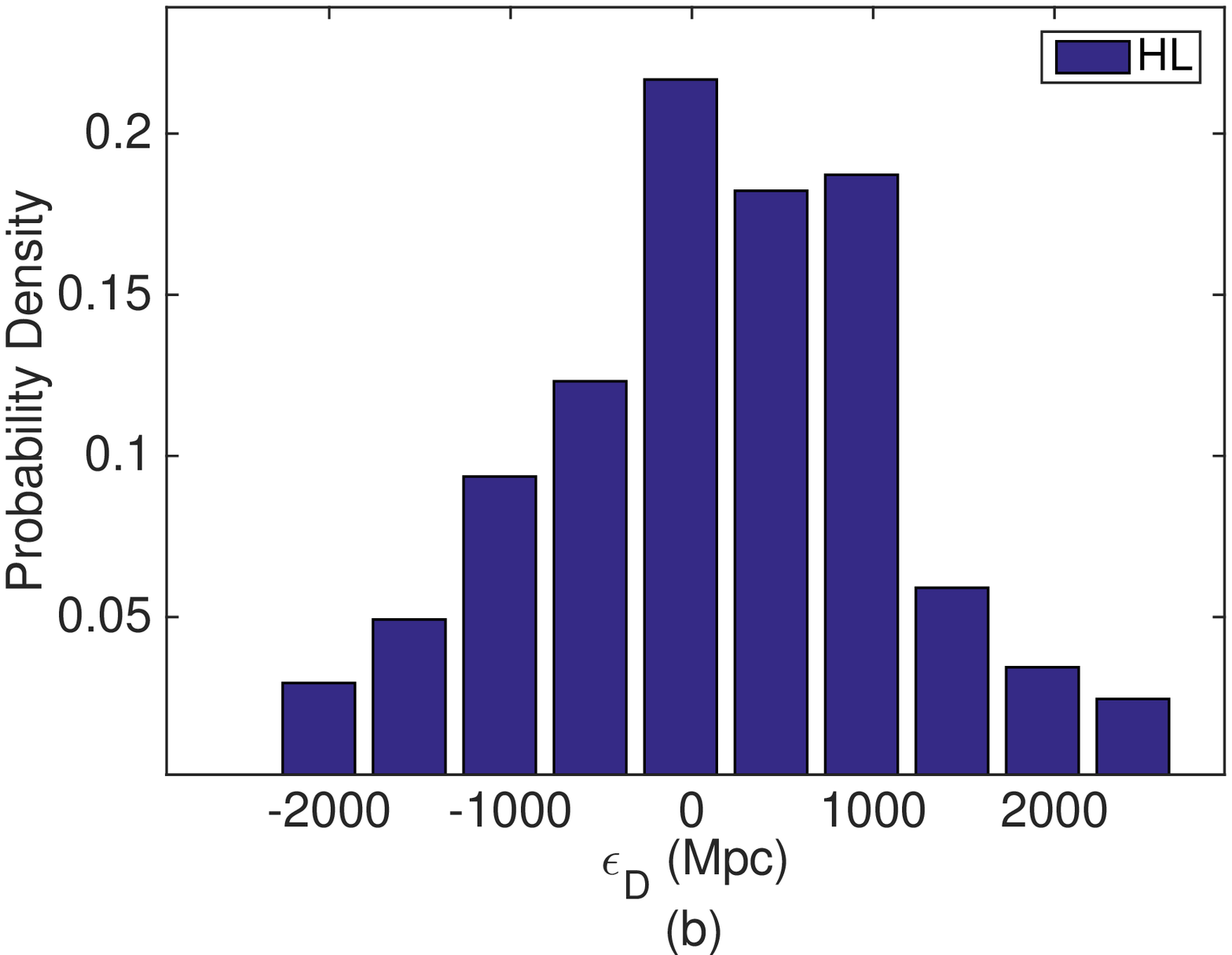}\includegraphics[height=4cm]{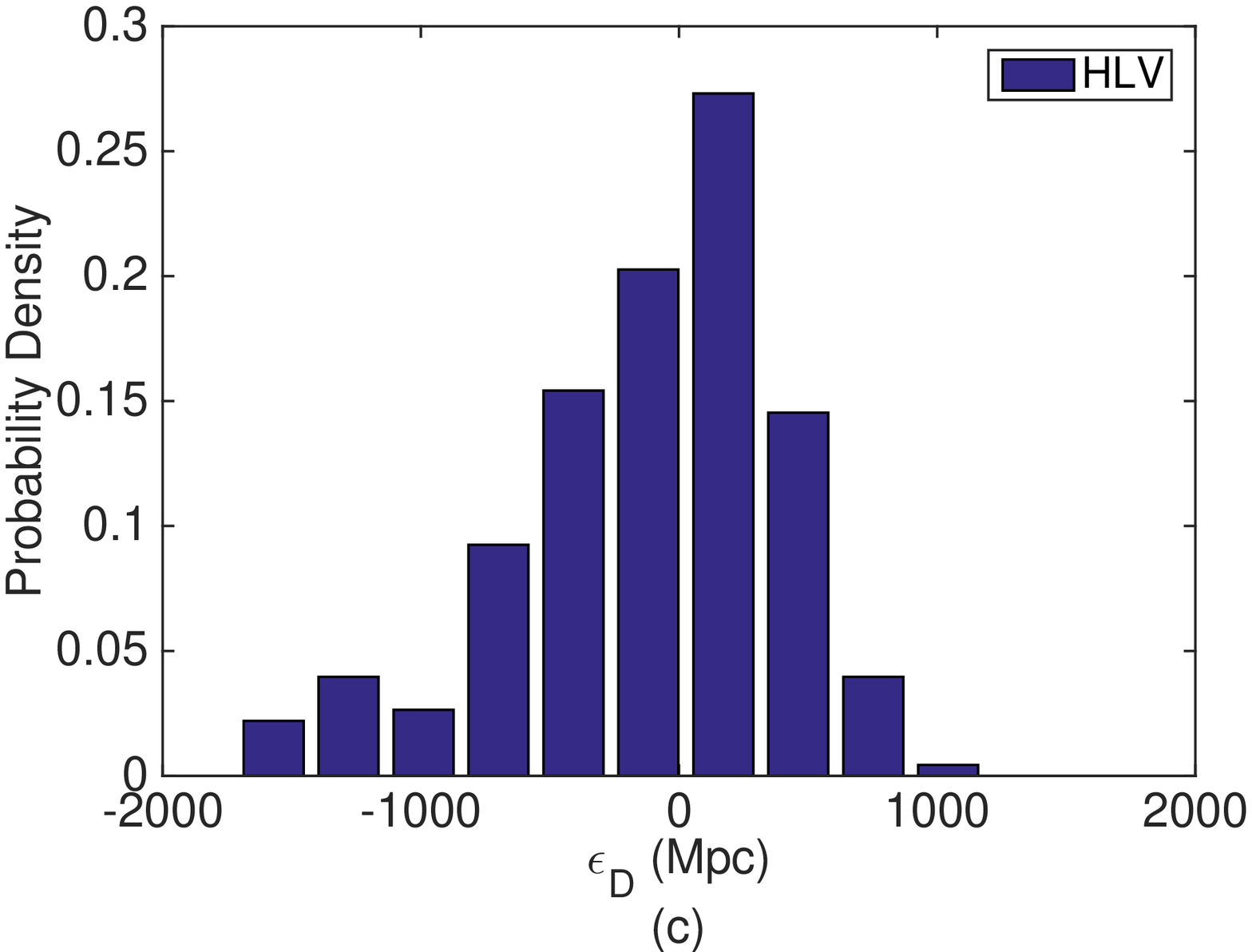}
\includegraphics[height=4cm]{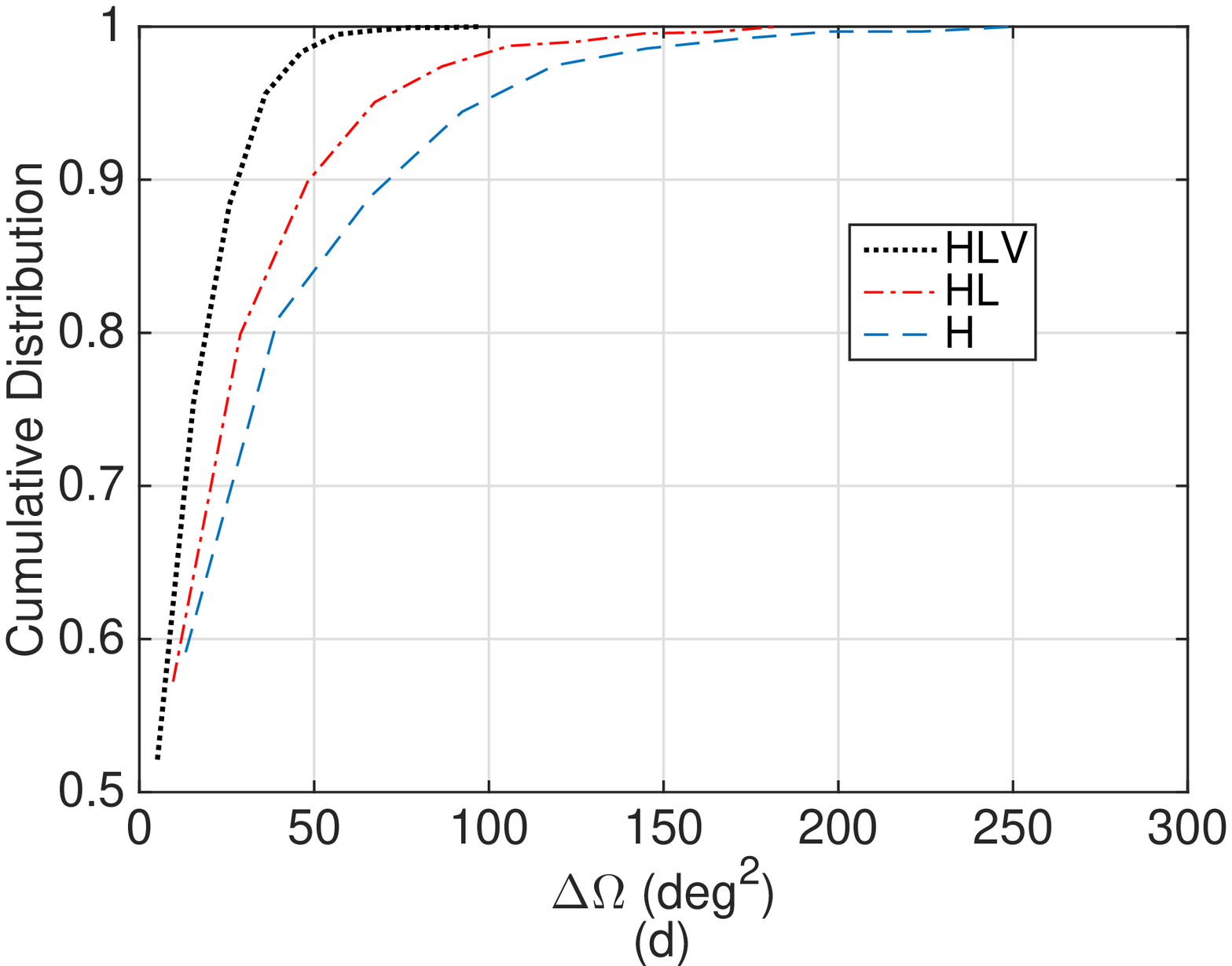}\includegraphics[height=4cm]{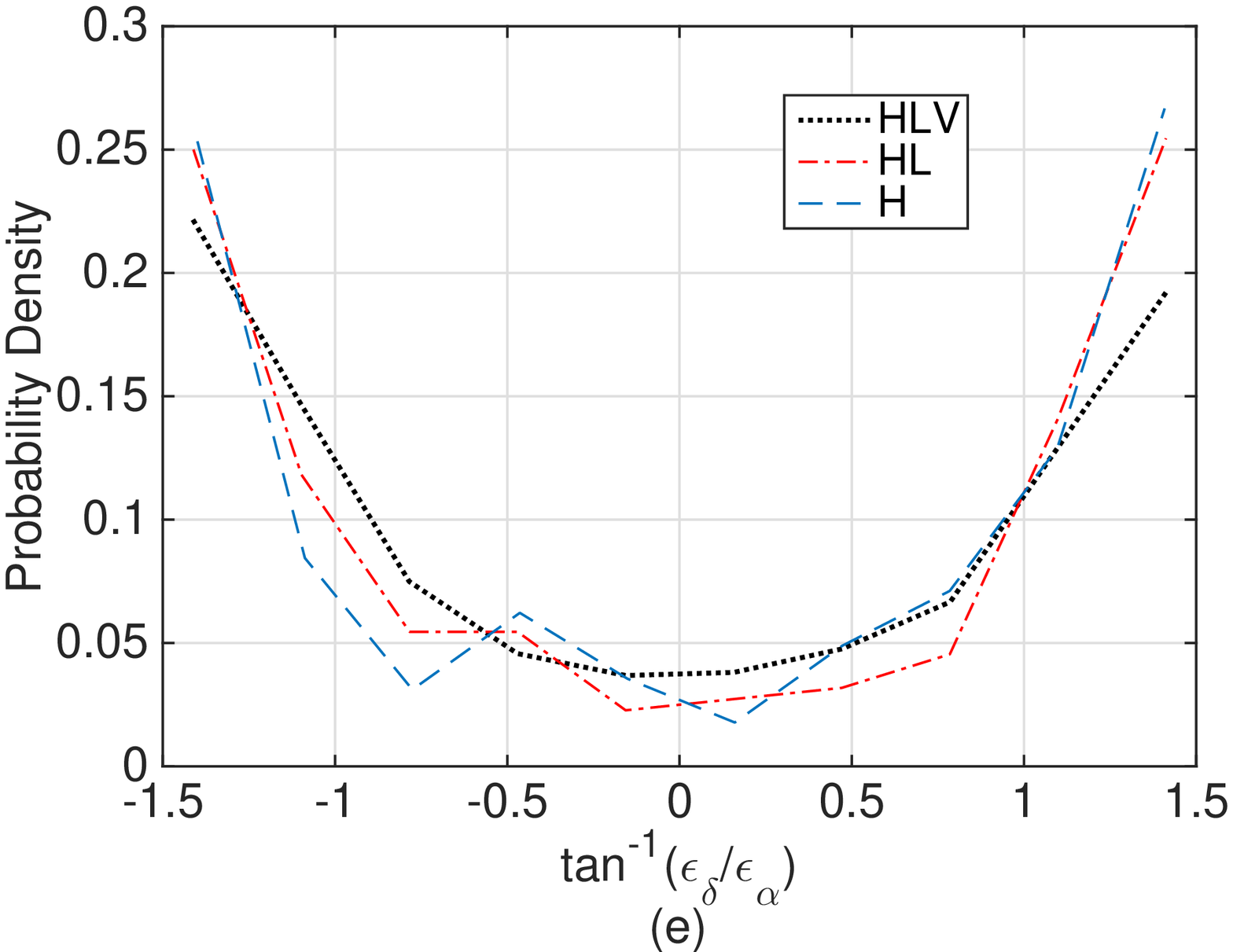}\includegraphics[height=4cm]{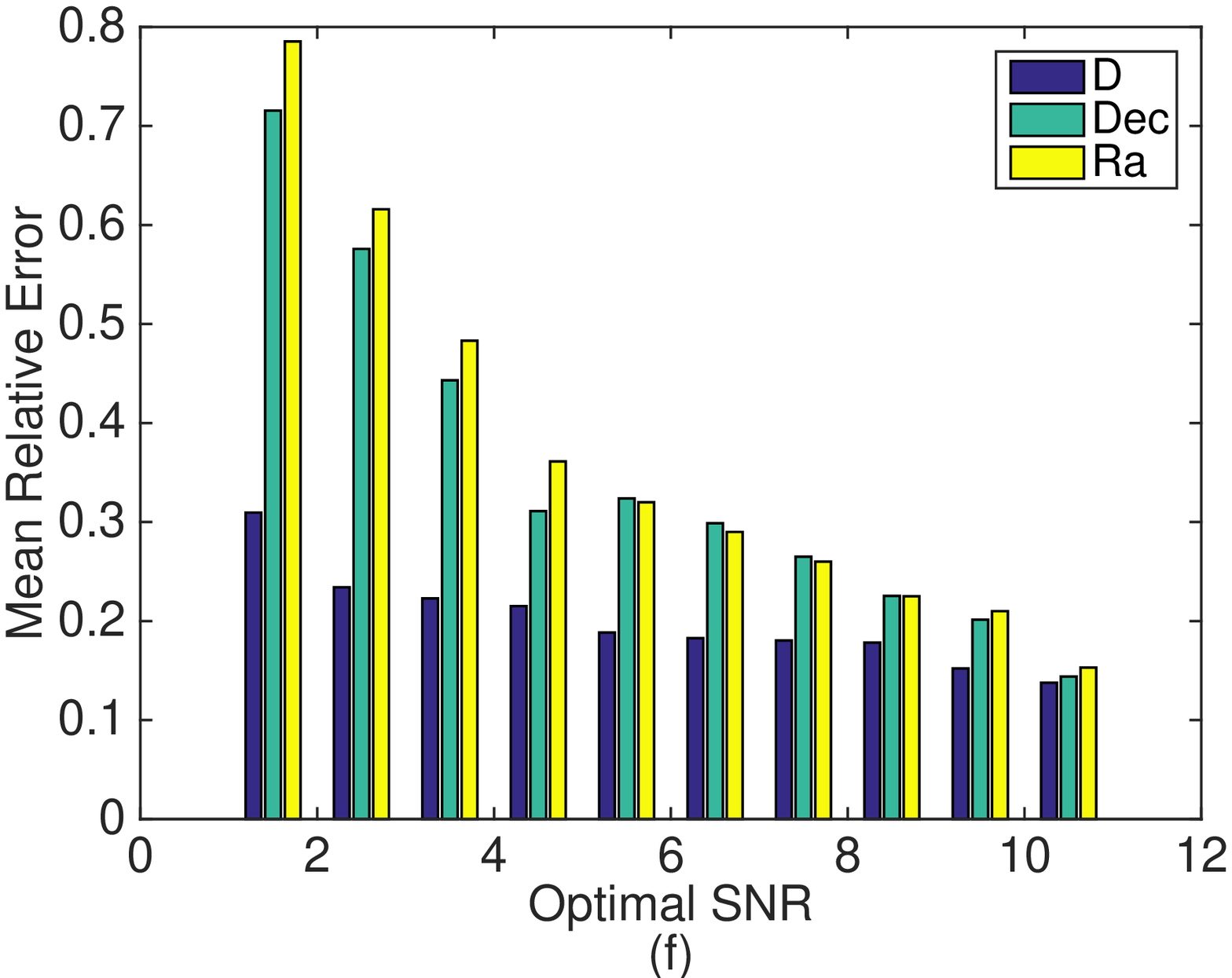}
\caption{ (a), (b), (c) Distribution of errors $\epsilon_{D}=y_{D}-Y_{D}$ in a multi-parameter predictor for advanced H, HL, and HLV, respectively. (d) Cumulative distribution of areas of angular parameter ($\alpha, \delta$) uncertainty. (e) Distribution probability of the gradient angle for $\epsilon_{\delta}$ on $\epsilon_{\alpha}$. Here, SNR = 8. (f) Mean relative errors of $D, \alpha, \delta$ in the multi-parameter estimation varying with optimal SNR. }\label{fig:acy_predictor}
\end{figure}

{\bf Results:} Our predictor is capable of estimating multiple parameters from simulated time series data in network detectors. By comparing the estimation errors for one, two, and three network detectors (Fig. \ref{fig:acy_predictor}), we have found that luminosity distance errors $\epsilon_{D}=y_{D}-Y_{D}$ decrease for a higher number of network detectors. In addition, the areas of angular uncertainty of celestial coordinates are constrained from several hundred deg$^{2}$ for one detector to 100 deg$^{2}$ for three network detectors. Fig. \ref{fig:acy_predictor} (e) indicates that the gradient direction of the declination error, $\epsilon_{\delta}$, on the right ascension error, $\epsilon_{\alpha}$, tends to be vertical, which is in accordance with existing error ellipses of angular parameters of all-sky maps \cite{angularresolution}. Furthermore, when SNR $\geq8$, the relative errors of $D$,$\alpha$,and $\delta$ in a multi-parameter predictor for advanced LIGO and advanced Virgo are all less than 23\% for three network detectors. 

\begin{figure}
\includegraphics[height=5cm]{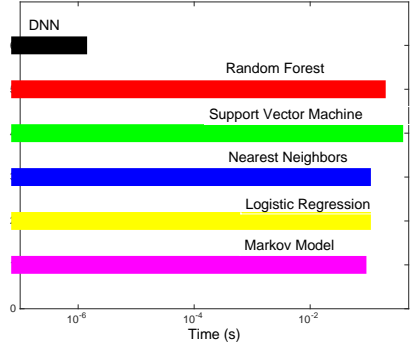}
 \caption{Data processing times for classification and prediction by our DNN and by other machine learning methods. Note: times refer to the processing of one test set by our DNN (using GPU:NVIDIA GPU (Gtx1080ti, 11G ram)) and the processing of one test set by other machine learning methods (using CPU: 2.7 GHz Intel Core i5)}\label{fig:Testingtime}
\end{figure}

\section{Conclusion}
We have investigated the robustness of DNNs in coincident GW searches and multiple space parameter estimation in advanced LIGO and advanced Virgo, and have also shown that Deep Filtering can be extended to the network detection of binary black hole mergers. Based on the features of a compact binary star's GW waveform, 100,000 theoretical inspiral-merger-ringdown signals are simulated to train our DNN. The training process takes approximately nine hours, after which the instant classification or prediction with a specific SNR is achieved (shown in. Fig.\ref{fig:Testingtime}). From these results, we can determine the fundamental relation between the performance of the DNN and the optimal SNR of the data sets. This is meaningful for broadening the scope of current GW searches in advanced LIGO and advanced Virgo and for future real-time multi-messenger astrophysics. 

For the above DNN structure, we have discussed signal-to-noise classification and space parameter estimation in one, two, and three detectors. The signal-to-noise classification indicates that the DNN can correctly classify the GW signals of compact binary stars with 100$\%$ detection sensitivity for time series with a SNR $\geq9$. From the parameter estimation, the DNN is able to estimate space parameters simultaneously. Our results also confirm that advanced LIGO and advanced Virgo can better estimate luminosity distance and sky location with a greater number of network detectors. Moreover, when a deep learning algorithm is used in advanced LIGO's and advanced Virgo's pipeline, there will be powerful instruments and equipment such as FPGAs for large data processing, and the corresponding sensitivity will be greatly improved. At present, our work mainly provides inspiration for deep learning in the network detection and source location of advanced LIGO and advanced Virgo. 

Furthermore, the main advantage of adding a DNN to GW searches is that retraining a DNN is time-saving once the DNN is trained well at a given PSD of advanced LIGO and advanced Virgo \cite{MLGW2}. Using a DNN facilitates real-time coincident GW detection, since a trained DNN is very efficient in performing classification and prediction. Based on the instant time detection and space parameter estimation of binary black hole mergers, the real-time observation of binary neutron star coalescence will likely be determined in the near future, which is highly important for the search for electromagnetic counterparts \cite{add4,add5,add6}.

\section*{\bf Acknowledgements}
This work was supported by the National Natural Science Foundation of Chi- na (Grant Nos. 11873001, 11633001, 11673008, and 61501069), the Nat- ural Science Foundation of Chongqing (Grant No. cstc2018jcyjAX0767), the Strategic Priority Program of the Chinese Academy of Sciences (Grant No. XDB23040100), Newton International Fellowship Alumni Follow-on Funding and the Fundamental Research Funds for the Central Universi- ties Project (Grant Nos. 106112017CDJXFLX0014, and 106112016CD- JXY300002). Our work is also supported by Chinese State Scholarship Fund and Newton International Fellowship Alumni Follow on Funding. We would like to express our great gratitude to Dr. George and Huerta for their inspi- ration and patient help.



\begin{thebibliography}{nbound}
\bibitem{prlnew} B. P. Abbott, et al. (LIGO Scientific Collaboration and Virgo Collabo- ration), Phys. Rev. Lett. 119, 161101 (2017), arXiv: 1710.05832.
\bibitem{prl4} B. P. Abbott, et al. (LIGO Scientific Collaboration and Virgo Collabo- ration), Phys. Rev. Lett. 119, 141101 (2017), arXiv: 1709.09660.
\bibitem{prl1} B. P. Abbott, et al. (LIGO Scientific Collaboration and Virgo Collabo- ration), Phys. Rev. Lett. 116, 061102 (2016), arXiv: 1602.03837.
\bibitem{prl2} B. P. Abbott, et al. (LIGO Scientific Collaboration and Virgo Collabo- ration), Phys. Rev. Lett. 116, 241103 (2016), arXiv: 1606.04855.
\bibitem{prl3} B. P. Abbott, et al. (LIGO Scientific Collaboration and Virgo Collabo- ration), Phys. Rev. Lett. 118, 221101 (2017), arXiv: 1706.01812.
\bibitem{add3} J. Li, and X. L. Fan, Sci. China-Phys. Mech. Astron. 60, 120431 (2017).
\bibitem{AdLIGO} B. P. Abbott, et al. (LIGO Scientific Collaboration and Virgo Collabo- ration), Class. Quantum Grav. 32, 074001 (2015), arXiv: 1411.4547.
\bibitem{AdVirgo} F. Acernese, et al. (Virgo Collaboration), Class. Quantum Grav. 32, 024001 (2015), arXiv: 1408.3978.
\bibitem{add1} D. Blair, L. Ju, C. N. Zhao, L. Q. Wen, Q. Chu, Q. Fang, R. G. Cai, J. R. Gao, X. C. Lin, D. Liu, L. A. Wu, Z. H. Zhu, D. H. Reitze, K. Arai, F. Zhang, R. Flaminio, X. J. Zhu, G. Hobbs, R. N. Manchester, R. M. Shannon, C. Baccigalupi, W. Gao, P. Xu, X. Bian, Z. J. Cao, Z. J. Chang, P. Dong, X. F. Gong, S. L. Huang, P. Ju, Z. R. Luo, L. E. Qiang, W. L. Tang, X. Y. Wan, Y. Wang, S. N. Xu, Y. L. Zang, H. P. Zhang, Y. K. Lau, and W. T. Ni, Sci. China-Phys. Mech. Astron. 58, 120402 (2015), arXiv: 1602.02872.
\bibitem{add2} D.Blair,L.Ju,C.N.Zhao,L.Q.Wen,H.X.Miao,R.G.Cai,J.R. Gao, X. C. Lin, D. Liu, L. A. Wu, Z. H. Zhu, G. Hammond, H. J. Paik, V. Fafone, A. Rocchi, C. Blair, Y. Q. Ma, J. Y. Qin, and M. Page, Sci. China-Phys. Mech. Astron. 58, 120405 (2015), arXiv: 1602.05087.
\bibitem{MLGW1} R. Biswas, L. Blackburn, J. Cao, R. Essick, K. A. Hodge, E. Kat- savounidis, K. Kim, Y. M. Kim, E. O. Le Bigot, C. H. Lee, J. J. Oh, S. H. Oh, E. J. Son, Y. Tao, R. Vaulin, and X. Wang, Phys. Rev. D 88, 062003 (2013), arXiv: 1303.6984.
\bibitem{MLGW2} D. George, and E. A. Huerta, Phys. Rev. D 97, 044039 (2018), arXiv: 1701.00008.
\bibitem{MLGWadd} D. George and E. A. Huerta, Phys. Lett. B, 778 (2018) arXiv:1711.03121.
\bibitem{MLGW3} A. Mytidis, A. A. Panagopoulos, O. P. Panagopoulos, B. Whiting, 2015, arXiv: 1508.02064 (2015).
\bibitem{MLGW4} A. Torres-Forne, A. Marquina, J. A. Font, and J. M. Ibanez, Phys. Rev. D 94, 124040 (2016), arXiv: 1612.01305.
\bibitem{MLGW5} K. A. Hodge, The Search for Gravitational Waves from the Coalescence of Black Hole Binary Systems in Data from the LIGO and Virgo Detectors Or: A Dark Walk through a Random Forest, PhD thesis, Dissertation for the Doctoral Degree, (California Institute of Technology, Pasadena, 2014).
\bibitem{new17} P. T. Baker, S. Caudill, K. A. Hodge, D. Talukder, C. Capano, and N. J. Cornish, Phys. Rev. D 91, 062004 (2015), arXiv: 1412.6479.
\bibitem{new18}  T. Gebhard, and N. Kilbertus and G. Parascandolo and I. Harry and B. Scholkopf, in Workshop on Deep Learning for Physical Sciences (DLPS) at the 31st Conference on Neural Information Processing Sys- tems (NIPS), Long Beach, CA, USA, 2017.
\bibitem{new19} H. Gabbard, M. Williams, F. Hayes, and C. Messenger, Phys. Rev. Lett. 120, 141103 (2018), arXiv: 1712.06041.
\bibitem{new20} X. R. Li, W. L. Yu, X. L. Fan, 2017, arXiv: 1712.00356.

\bibitem{MLoverview} J. G Carbonell, R. S. Michalski, and T. M. Mitchell. An overview of machine learning. In Machine learning, pages 3$-$23. Springer Berlin Heidelberg, Heidelberg, Germany (1983).
\bibitem{rep17} W. S. McCulloch and W. Pitts, Bull. Math. Biophys., 5(4):115$-$133 (1943).
\bibitem{rep18} G. A Carpenter, Neural networks, 2(4):243$-$257 (1989).
\bibitem{DNN1} J. Schmidhuber, Neural Networks 61 85 (2015).
\bibitem{RNN1} J. Wang, SIAM J. Sci. Comput., 18(5), 1479$-$1493 (2006).
\bibitem{UIUC66} K. Fukushima, Biological Cybernetics 36, 193 (1980).
\bibitem{UIUC67} Y. LeCun and Y. Bengio (MIT Press, Cambridge, MA, USA) Chap. Convolutional Networks for Images, Speech, and Time Series, pp. 255$-$258 (1998).
\bibitem{UIUC68} A. Krizhevsky, I. Sutskever, and G. E. Hinton, in Advances in Neural Information Processing Systems 25, edited by F. Pereira, C. J. C. Burges, L. Bottou, and K. Q. Weinberger (Curran Associates, Inc. Nice, France) pp. 1097$-$1105 (2012).
\bibitem{BNS1} L. Li and B. Paczy$\acute{n}$ski, ApJ, 507: 59 (1998).
\bibitem{BNS2} B. D. Metzger and E. Berger, ApJ, 746: 48 (2012).
\bibitem{Mmess1} B.P. Abbott et al. (LIGO Scientific Collaboration and Virgo Collaboration), Astrophys. J. Lett. 848, L12 (2017).
\bibitem{Mmess2} D. A. Coulter et al., GCN 21529, 1, 1105 Media, Inc. McLean, VA  (2017).
\bibitem{Mmess3} D.A. Coulter et al., Science, in press (2017), DOI: 10.1126/science.aap9811.
\bibitem{Mmess4} LIGO Scientific Collaboration and Virgo Collaboration, GCN 21513, 1,1105 Media, Inc. McLean, VA (2017).
\bibitem{Fan} X. L. Fan, Sci. China-Phys. Mech. Astron. 59, 640001 (2016).
\bibitem{addfan} C. Cutler, E. E. Flanagan, Phys. Rev. D 49, 2658 (1994).
\bibitem{detectorlocation} P. Jaranowski et al., Phys. Rev. D 58, 063001 (1998). arXiv:gr-qc/9804014v1.
\bibitem{HLsensitivity} D. Shoemaker, Advanced LIGO anticipated sensitivity curves-LIGO Document (2010).
\bibitem{Vsensitivity} F. Acernese, M. Agathos, K. Agatsuma, D. Aisa, N. Allemandou, A.
Allocca, J. Amarni, P. Astone, G. Balestri, G. Ballardin, F. Barone, J. P. Baronick, M. Barsuglia, A. Basti, F. Basti, T. S. Bauer, V. Bavigadda, M. Bejger, M. G. Beker, C. Belczynski, D. Bersanetti, A. Bertolini, M. Bitossi, M. A. Bizouard, S. Bloemen, M. Blom, M. Boer, G. Bogaert, D. Bondi, F. Bondu, L. Bonelli, R. Bonnand, V. Boschi, L. Bosi, T. Bouedo, C. Bradaschia, M. Branchesi, T. Briant, A. Brillet, V. Brisson, T. Bulik, H. J. Bulten, D. Buskulic, C. Buy, G. Cagnoli, E. Calloni, C. Campeggi, B. Canuel, F. Carbognani, F. Cavalier, R. Cavalieri, G. Cella, E. Cesarini, E. C. Mottin, A. Chincarini, A. Chiummo, S. Chua, F. Cleva, E. Coccia, P. F. Cohadon, A. Colla, M. Colombini, A. Conte, J. P. Coulon, E. Cuoco, A. Dalmaz, S. DAntonio, V. Dattilo, M. Davier, R. Day, G. Debreczeni, J. Degallaix, S. Deleglise, W. D. Pozzo, H. Dereli, R. D. Rosa, L. D. Fiore, A. D. Lieto, A. D. Virgilio, M. Doets, V. Dolique, M. Drago, M. Ducrot, G. Endroczi, V. Fafone, S. Farinon, I. Ferrante, F. Ferrini, F. Fidecaro, I. Fiori, R. Flaminio, J. D. Fournier, S. Franco, S. Frasca, F. Frasconi, L. Gammaitoni, F. Garufi, M. Gaspard, A. Gatto, G. Gemme, B. Gendre, E. Genin, A. Gennai, S. Ghosh, L. Giacobone, A. Giazotto, R. Gouaty, M. Granata, G. Greco, P. Groot, G. M. Guidi, J. Harms, A. Heidmann, H. Heitmann, P. Hello, G. Hemming, E. Hennes, D. Hofman, P. Jaranowski, R. J. G. Jonker, M. Kasprzack, F. Kefelian, I. Kowalska, M. Kraan, A. Krolak, A. Kutynia, C. Lazzaro, M. Leonardi, N. Leroy, N. Letendre, T. G. F. Li, B. Lieunard, M. Lorenzini, V. Loriette, G. Losurdo, C. Magazzu, E. Majorana, I. Maksimovic, V. Malvezzi, N. Man, V. Mangano, M. Mantovani, F. Marchesoni, F. Marion, J. Marque, F. Martelli, L. Martellini, A. Masserot, D. Meacher, J. Meidam, F. Mezzani, C. Michel, L. Milano, Y. Minenkov, A. Moggi, M. Mohan, M. Montani, N. Morgado, B. Mours, F. Mul, M. F. Nagy, I. Nardecchia, L. Naticchioni, G. Nelemans, I. Neri, M. Neri, F. Nocera, E. Pacaud, C. Palomba, F. Paoletti, A. Paoli, A. Pasqualetti, R. Passaquieti, D. Passuello, M. Perciballi, S. Petit, M. Pichot, F. Piergiovanni, G. Pillant, A. Piluso, L. Pinard, R. Poggiani, M. Prijatelj, G. A. Prodi, M. Punturo, P. Puppo, D. S. Rabeling, I. Racz, P. Rapagnani, M. Razzano, V. Re, T. Regimbau, F. Ricci, F. Robinet, A. Rocchi, L. Rolland, R. Romano, D. Rosinska, P. Ruggi, E. Saracco, B. Sassolas, F. Schimmel, D. Sentenac, V. Sequino, S. Shah, K. Siellez, N. Straniero, B. Swinkels, M. Tacca, M. Tonelli, F. Travasso, M. Turconi, G. Vajente, N. van Bakel, M. van Beuzekom, J. F. J. van den Brand, C. Van Den Broeck, M. V. van der Sluys, J. van Heijningen, M. Vasuth, G. Vedovato, J. Veitch, D. Verkindt, F. Vetrano, A. Vicere, J. Y. Vinet, G. Visser, H. Vocca, R. Ward, M. Was, L. W. Wei, M. Yvert, A. Z. zny, and J. P. Zendri, Class. Quantum Grav. 32, 024001 (2015), arXiv: 1408.3978.
\bibitem{networksGWreviewlatest} Bernard F Schutz, Classical Quant. Grav., 28, 12 (2011).
\bibitem{GWrevpaper} B.S. Sathyaprakash, B.F. Schutz, Living Rev. Relativity 12, 2 (2009). arXiv: 0903.0338v1.
\bibitem{wavelet} B. P. Abbott et al. (LIGO Scientific Collaboration and Virgo Collaboration), Phys. Rev. Lett. 116, 241102 (2016).
\bibitem{CNN1} K. O'Shea, R. Nash, arXiv:1511.08458 (2015).
\bibitem{cv} D. Mishkin, N. Sergievskiy, and J. Matas, Comput. Vision Image Underst. 161, 11 (2017).
\bibitem{lp} R. Collobert, J. Weston, L. Bottou, M. Karlen, K. Kavukcuoglu, P. Kuksa, J. Mach. Learn. Res.,12, 2493-2537 (2011).
\bibitem{timeCNN} Y. Zheng et al., Time Series Classification Using Multi-Channels Deep Convolutional Neural Networks, Web-Age Information Management pp 298$-$310, Springer, Cham, Switzerland (2014).
\bibitem{UIUC102} J. Snoek, H. Larochelle, and R. P. Adams, in Advances in Neural Information Processing Systems 25, edited by F. Pereira, C. J. C. Burges, L. Bottou, and K. Q. Weinberger (Curran Associates, Inc.) pp. 2951$-$2959 (2012).
\bibitem{GenCNN} L.X. Xie, A. Yuille, arXiv:1703.01513v1(2017).
\bibitem{ADAM} Diederik P. Kingma, Jimmy Ba, arXiv:1412.6980(2017).
\bibitem{rep24} D. E. Rumelhart, G. E. Hinton,  R. J. Williams, Learning representations by back-propagating errors, Cognitive modeling 5(3):1 (1988).
\bibitem{BP1} J. Li, J. Cheng, J. Shi, F. Huang, Brief introduction of back propagation (BP) neural network algorithm and its improvement, Advances in Computer Science and Information Engineering, (Springer, Berlin, 2012) pp 553$-$558.
\bibitem{angularresolution} L. Q. Wen, Y. B Chen, Geometrical expression for the angular resolution of a network of gravitational-wave detectors, Phys. Rev. D 81, 082001 (2010).
\bibitem{add4} Z. J. Cao, Sci. China-Phys. Mech. Astron. 59, 110431 (2016).
\bibitem{add5} H. Gao, Sci. China-Phys. Mech. Astron. 61, 059531 (2018).
\bibitem{add6} T.P.Li,S.L.Xiong, S.N.Zhang, F.J.Lu, L.M.Song, X.L.Cao, Z. Chang, G. Chen, L. Chen, T. X. Chen, Y. Chen, Y. B. Chen, Y. P. Chen, W.Cui, W.W.Cui, J.K.Deng, Y.W.Dong,Y.Y.Du, M.X.Fu, G.H.Gao, H.Gao, M.Gao, M.Y.Ge, Y.D.Gu, J.Guan, C.C.Guo, D.W. Han, W. Hu, Y. Huang, J. Huo, S. M. Jia, L. H. Jiang, W. C. Jiang, J. Jin, Y.J.Jin, B.Li, C.K.Li, G.Li, M.S.Li, W.Li, X.Li, X.B.Li, X.F.Li, Y.G.Li, Z.J.Li, Z.W.Li, X.H.Liang, J.Y.Liao, C.Z.Liu, G.Q.Liu, H.W.Liu, S.Z.Liu, X.J.Liu,Y.Liu, Y.N.Liu, B.Lu, X. F.Lu, T.Luo, X.Ma, B.Meng,Y.Nang, J.Y.Nie, G.Ou, J.L.Qu, N.Sai, L.Sun,Y.Tan, L.Tao, W.H.Tao, Y.L.Tuo, G.F.Wang, H. Y. Wang, J. Wang, W. S. Wang, Y. S. Wang, X. Y. Wen, B. B. Wu, M. Wu, G.C.Xiao, H.Xu, Y.P.Xu, L.L.Yan, J.W.Yang, S.Yang,Y.J. Yang, A. M. Zhang, C. L. Zhang, C. M. Zhang, F. Zhang, H. M. Zhang, J. Zhang, Q. Zhang, S. Zhang, T. Zhang, W. Zhang, W. C. Zhang, W. Z. Zhang, Y. Zhang, Y. Zhang, Y. F. Zhang, Y. J. Zhang, Z. Zhang, Z. L. Zhang, H. S. Zhao, J. L. Zhao, X. F. Zhao, S. J. Zheng, Y. Zhu, Y. X. Zhu, and C. L. Zou, Sci. China-Phys. Mech. Astron. 61, 031011 (2018), arXiv: 1710.06065.


\end{thebibliography}
\end{document}